\documentclass{article}
\usepackage{spconf,amsmath,graphicx}
\usepackage{multirow}
\usepackage{float}
\usepackage{dblfloatfix}
\usepackage{arydshln, multicol}
\usepackage{hyperref}
\usepackage{placeins}
\usepackage{booktabs}

\title{Multi-Channel MOSRA: Mean Opinion Score and Room Acoustics  Estimation Using Simulated Data and a Teacher Model}

\name{Jozef Coldenhoff$^{\text{ }1,2}\sthanks{jozef.coldenhoff@epfl.ch. This work was performed during an internship at Logitech.}$
Andrew Harper$^1$
Paul Kendrick$^1$
Tijana Stojkovic$^1$
Milos Cernak$^1$}
\address{$^1$Logitech Europe S.A., Lausanne, Switzerland \\
$^2$École Polytechnique Fédérale de Lausanne (EPFL), Lausanne, Switzerland \\}

\begin{document}
\ninept

\maketitle

\begin{abstract}
Previous methods for predicting room acoustic parameters and speech quality metrics have focused on the single-channel case, where room acoustics and Mean Opinion Score (MOS) are predicted for a single recording device. However, quality-based device selection for rooms with multiple recording devices may benefit from a multi-channel approach where the descriptive metrics are predicted for multiple devices in parallel. Following our hypothesis that a model may benefit from multi-channel training, we develop a multi-channel model for joint MOS and room acoustics prediction (MOSRA) for five channels in parallel. The lack of multi-channel audio data with ground truth labels necessitated the creation of simulated data using an acoustic simulator with room acoustic labels extracted from the generated impulse responses and labels for MOS generated in a student-teacher setup using a wav2vec2-based MOS prediction model. Our experiments show that the multi-channel model improves the prediction of the direct-to-reverberation ratio, clarity, and speech transmission index over the single-channel model with roughly 5$\times$ less computation while suffering minimal losses in the performance of the other metrics. 
\end{abstract}


\section{Introduction}
\label{sec:intro}

The Mean Opinion Score (MOS) is one of the simplest yet effective metrics of subjective speech quality and can be evaluated using a subjective listening test based on the ITU-T P.800 recommendation \cite{p800rec}, or its crowdsourcing approach described in the ITU-T P.808~\cite{p808rec}. 

However, given the costly nature of carrying out subjective listening tests, many methods have been developed to estimate this subjective metric for unevaluated speech recordings blindly. Most methods are based on neural networks trained to map speech to MOS scores. Nevertheless, many of these systems lack robustness due to the lack of large human-labeled datasets. Therefore, the ConferencingSpeech 2022 challenge \cite{yi2022conferencingspeech} recently released a larger dataset consisting of 86k audio files with crowdsourced MOS labels, with top performers in the challenge achieving a Pearson correlation of roughly 0.8  with human labels. 

An overall speech quality metric is very useful but does not provide insights into the causes of degraded speech quality. To counteract this, multi-valued non-intrusive speech quality assessment (NISQA) methods have been proposed \cite{Mittag_2021, reddy2022dnsmos}. Characterization of the listening environment also provides such insights. Therefore, efforts have been made in the blind estimation of room acoustic descriptors, where deep neural networks (DNNs) estimate a single acoustic descriptor, such as reverberation time measured in seconds, T60~\cite{t60estimation}, or the speech transmission index \cite{xiang23_interspeech}. Besides, joint prediction of more acoustic parameters using recurrent neural networks can estimate reverberation time (T60), clarity metrics in decibels (C50 and C80), direct to reverberant ratio (DRR) also in decibels~\cite{callens2020joint}. Finally, Lopez et al. \cite{Lopez_2021} further expanded the joint prediction by adding the signal-to-noise ratio (SNR, in decibels) and the speech transmission index (STI, ranging from 0 to 1). A lightweight model that jointly predicted MOS and room acoustic parameters was presented in \cite{hajal2022efficient}. Later, Hajal et al. investigated self-supervised learning methods for the same joint prediction and achieved state-of-the-art results~\cite{hajal2022mosra}. More recently, Sarabia et al. \cite{sarabia23_interspeech} released a corpus of simulated audio data with corresponding acoustic parameters where the authors showed that models trained on simulated data were able to generalize to real-world conditions. 

\begin{figure*}
    \centering
    \includegraphics[width=0.8\textwidth]{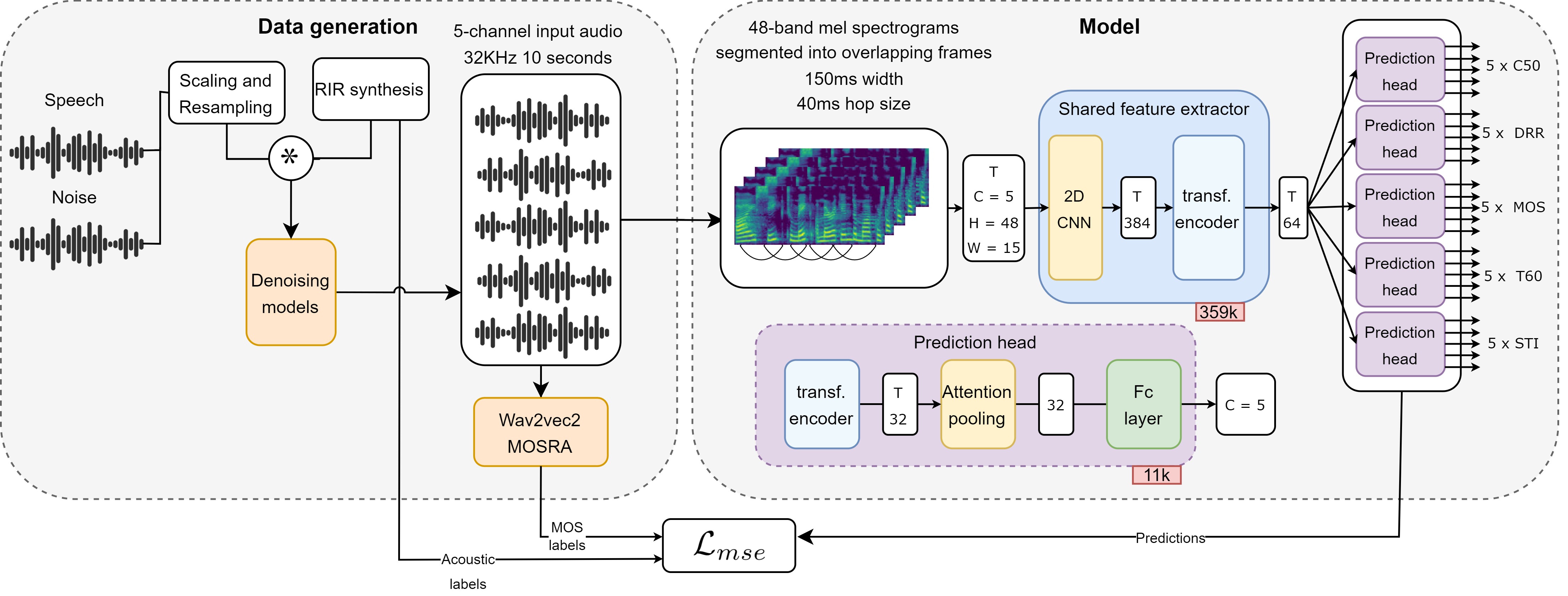}
    \caption{Overview of the proposed framework. On the left, a high-level overview of the data generation process is given. On the right, the details of the model architecture are shown, with the number of parameters shown in the red boxes.}
    \label{fig:framework}
\end{figure*}

The recent work on speech quality assessment and room acoustics estimation has focused on the single-device case where descriptive metrics are predicted for a single recording device in a one-to-one fashion. However, there has recently been interest in a many-to-one setup where multiple input channels map to a single metric. One such setting is the clarity prediction challenge \cite{cpc1}, where the challenge entrants are tasked with predicting the intelligibility of binaural hearing aid recordings achieved by human listeners.

Another application of a many-to-one setup is the task of device selection. The rise of smart home devices and personal digital devices led to the situation where multiple spatially disjoint microphones can simultaneously record a speaker. This naturally leads to the challenge of selecting which audio stream to transmit. Thus, in previous work on device selection in the context of smart home devices, Barber et al.~\cite{barber2022endtoend} designed a multi-channel system to predict the closest device to the current speaker, where an improvement was found over a signal processing baseline. Similarly, Yoshioka et al.~\cite{yoshioka2022picknet} presented PickNet, a system to predict the closest device, where they also note an improvement in word error rate (WER) when using a multi-channel model over their single-channel baseline. 

As noted, these previous works use the heuristic that the best device is the closest to the current speaker. However, choosing the closest device may not be optimal regarding audio quality, as many factors besides distance can influence the quality of an audio stream. For example, the closest device may be placed in a region with many reflections, decreasing DRR and clarity. It may also be placed next to a noise source or run a less powerful denoising model than other devices in the room.

Inspired by these previous works we aim to develop a system for quality-based device selection, where we extend our task setup to a many-to-many case. Here the model is tasked to predict multiple sets of quality and acoustic metrics given multiple input channels. Specifically, we aim to train a model that predicts speech quality MOS and a set of acoustic parameters (STI, T60, DRR, and C50) for five channels in parallel. Our motivation for this is two-fold. Firstly, having access to MOS and room acoustics will improve the interpretability of the device selection, thus aiding in making informed decisions. Secondly, we conjecture that a model trained to predict the metrics of interest may benefit from multi-channel training, as information contained in the different audio streams may give the model global information about the overall acoustic environment.

Thus, in this paper, we present multi-channel MOSRA, a model that predicts the MOS and room acoustics for five channels in parallel. Given the lack of multi-device data with ground truth labels, we simulate it using an acoustic simulator where the ground truth room acoustic labels are extracted from the room impulse responses, while the labels for MOS are generated by student-teacher setup using a wav2vec2 (XLSR)-based model trained to predict MOS scores \cite{hajal2022efficient}. Our experiments show that the multi-channel model improves the prediction of DRR, C50, and STI over the single-channel model with roughly 5$\times$ less computation while suffering minimal losses in the performance of the other metrics. 

\section{Methods}

This section proposes an extension to the MOSRA model that allows it to generate predictions for five channels in parallel. The overall model architecture is shown on the right in Fig. \ref{fig:framework}, and has a total number of 411k trainable parameters. Moreover, the used data simulation pipeline is described in detail.

\subsection{Multi-channel feature extractor}

The multi-channel feature extractor is an adapted version of the one used in \cite{hajal2022mosra}, where the first convolutional neural network (CNN) layer is modified to take as input five mel-spectrogram segments instead of a single one, with the rest of the network left unchanged.

These multi-channel Mel-spectrogram segments are obtained by Mel transforming the STFT representations of the five audio files. We use 48 Mel bands with an FFT window size of 20ms and a hop size of 10ms. Super wideband audio is supported as the maximum Mel band frequency is 16kHz. The resulting Mel-spectrogram is further divided into overlapping frames, each spanning 150ms with a hop size of 40ms.

Consequently, the Mel-spectrogram segments are passed to the CNN, which generates an embedding per segment. These multi-channel embeddings are then further processed by a transformer encoder \cite{hajal2022mosra}, which accounts for temporal dependencies.

\subsection{Multi-channel prediction heads}

Similarly to the feature extractor, we adapt the original prediction heads to output a metric estimate per channel. This is done by changing the final fully connected layer to have an output size of 5 instead of 1.

\subsection{Multi-channel meeting data generation}

Given the lack of suitable training data, we leverage acoustic simulation to generate it. An overview of the proposed simulation pipeline is shown on the left in Fig. \ref{fig:framework}. PyRoomAcoustics \cite{pyroomacoustics} is used to generate the room impulse responses (RIRs) which are convolved with the speech and noise sources. We use a combination of the image-source method (ISM), and ray tracing to generate the RIRs, where the maximum order for the ISM is set to 3 and the number of rays is computed automatically following Vorländer 2008, Eq. (11.12) \cite{raytracing}.

Then, for every unique scenario, we randomly generate a shoebox-shaped room between 2.1-10m in width/length and 2-4m in height. The absorption coefficients of the room materials are randomized to reflect a realistic acoustic environment. This results in a mean T60 of 0.41s with a standard deviation of 0.18. We then place a speech source at a height between 1.3-2m at least 10cm from a wall. Then, one or two noises are added to the room in a random location with the constraint that they are at least 10cm from any wall. Finally, each of the five microphones is placed in one of two configurations, either being mounted on a wall or in approximately the center of the room, reflecting a device placed on a table.

We use clean speech from the LibriSpeech train-clean-100, dev-clean, test-clean sets \cite{Pratap2020MLSAL} as source data. The noise data is taken from the DNS challenge \cite{dubey2022icassp}, which we randomly partitioned in train, validation, and test splits at fractions of 0.8, 0.1, and 0.1, leading to 43902, 5487, 5489 noise files per subset respectively.

Thus, after generating the room impulse responses, we randomly choose a clean speech file and up to two noise files and resample them from 16kHz to 32kHz. The noise and speech samples are then repeated to have a length of 10 seconds. Then, the dB full scale (dBFS) levels of the speech and noise are sampled as defined in Eq.~\ref{eq:dBFS}, which is done to obtain a desirable distribution of SNR at the outputs.

\begin{equation}
 \label{eq:dBFS}
  \begin{split}
    \text{dBFS(Noise)} \sim (Beta(1.5, 1.5) \cdot -40) -20 \\
    \text{dBFS(Speech)} \sim (Beta(1.5, 1.5) \cdot -30) -10
  \end{split}
\end{equation}

The scaled audio segments are then convolved with the RIRs to obtain the five output mixtures. These are then uniformly scaled such that the loudest signal in the mixtures has a dBFS between -20 and 0. To obtain the final audio, the wet mixtures are passed through three different internal Logitech denoising models to simulate the case where the recording devices apply denoising before transmitting the signal to a central hub.

To obtain the labels, we compute the room acoustic parameters from the RIRs: STI \cite{houtgast1971evaluation}, DRR \cite{ACE-challenge}, T60 and C50 \cite{ISO3382}, . Since we do not have access to MOS labels for our generated data, we leverage a student-teacher setup. Consequently, the audio outputs are passed to a larger single-channel network trained to estimate MOS. We used an XLSR-based model trained to predict MOS scores on single-channel recordings as our teacher model \cite{hajal2022efficient}. 

In the end we obtain three multi-channel datasets with a size in hours of 79.3, 7.5, and 6.3 for training, validation, and testing, respectively. The length of every individual audio clip is slightly longer than 10 seconds due to the convolution with the RIRs. Finally, the simulation resulted in the distributions of T60 and DRR versus distance to the source, shown in Fig. \ref{fig:T60_DRR}. The figure highlights that distance is only loosely correlated with acoustic parameters.

\subsection{Loss function}

To train the network, we use the mean squared error (MSE) with the same loss weights as in \cite{hajal2022mosra}. The loss used for training is shown in Eq. \ref{eq:loss}. Where $N$ is the number of samples in a minibatch.

\begin{equation}
 \label{eq:loss}
    \begin{split}
        \text{MSE} &= \frac{1}{N} \Sigma_1^N (y_{pred_n} - y_{true_n})^2 \\   
         \mathcal{L}_{\text{Room acoustics}} &= \text{MSE}_{T60} + \text{MSE}_{C50} + \text{MSE}_{STI} + \text{MSE}_{DRR} \\
       \mathcal{L}_{overall} &= 2 \cdot \text{MSE}_{\text{MOS}} + 0.2 \cdot \mathcal{L}_{\text{Room acoustics}}    
    \end{split}          
\end{equation}

Similarly to the original training scheme, we normalize the labels with the mean and standard deviations across the dataset to ensure that all the model outputs have the same range.

\section{Experimental setup}

\begin{figure}
    \centering
    \includegraphics[width=0.4\textwidth]{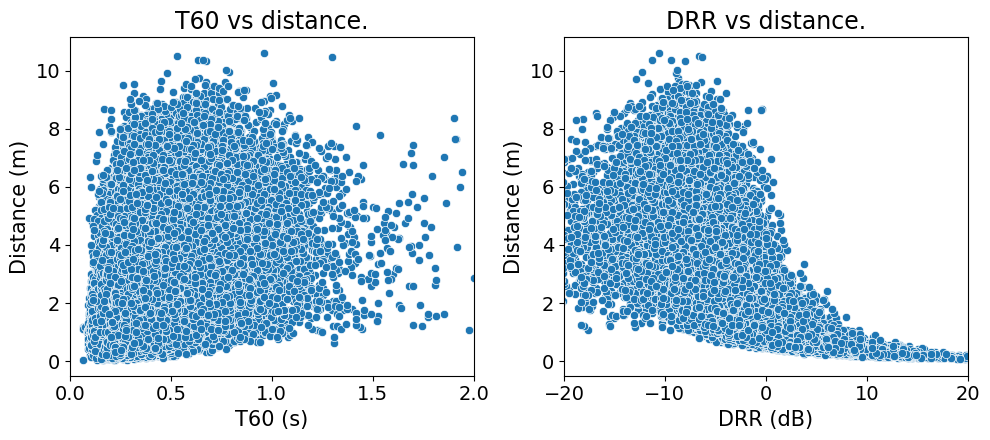}
    \caption{T60 and DRR versus distance to the active speaker.}
    \label{fig:T60_DRR}
\end{figure}

\subsection{Baseline system}

To show the benefit of multi-channel training and inference, we compare the proposed model against a baseline. The chosen baseline model is the original single-channel MOSRA model \cite{hajal2022mosra} trained on the individual channels of our simulated data. Both the single and multi-channel models have a similar number of parameters with 411k and 413K parameters, respectively. 

\subsection{Training and evaluation}

Both models were trained using the ADAM \cite{kingma2017adam} optimizer, a batch size of 32, a learning rate of 5e-4, and early stopping with patience of 30 on the Pearson correlation of the MOS predictions. Furthermore, the learning rate was decreased by a factor of 10 if the validation loss did not decrease for 15 epochs. Note that the single-channel baseline model had 5$\times$ more training steps per epoch, one for each channel per unique simulated room.

During multi-channel model training, we sample five devices from a room with replacement to simulate the case that less than five microphones are present i.e. some channels are repeated in the input to the model. This also ensures that the model does not learn any patterns from the ordering of the channels. Note that during testing, all channels in the room are passed to the model. For evaluation purposes, RMSE will be used for the acoustic parameters (STI, T60, DRR, and C50). For MOS, the Pearson correlation metric is used.

\section{Results}

\subsection{Acoustic parameters}

The results for the prediction of acoustic parameters are shown in Table \ref{table:ACOUSTIC_RES}. We can observe that the multi-channel system outperforms the baseline single-channel system in STI, DRR, and C50. The confidence intervals also show that the improvement in the three acoustic parameters is significant.

\begin{table}
\begin{tabular}{lllll}
\toprule
Model       & Multi-channel & Single-channel \\
            & RMSE          & (baseline) RMSE                     \\
\midrule
    STI     & \textbf{0.017} [0.0170, 0.0176] & 0.018 [0.0178, 0.0184]  \\
    DRR [dB] & \textbf{2.78} [2.736, 2.839] & 3.23 [3.190, 3.267] \\
    T60 [s] & 0.11 [0.105, 0.112] & \textbf{0.09} [0.090, 0.097]\\
    C50 [dB] & \textbf{1.80} [1.757, 1.834] & 1.87 [1.835, 1.912]\\
    
\bottomrule
\end{tabular}
\caption{RMSE ($\downarrow$) of the baseline and multi-channel models on acoustic parameters using the simulated test set. The confidence intervals are obtained by bootstrapping with 1000 repetitions.}
\label{table:ACOUSTIC_RES}
\end{table}

\begin{figure*}
\centering
    \includegraphics[width=0.75\textwidth]{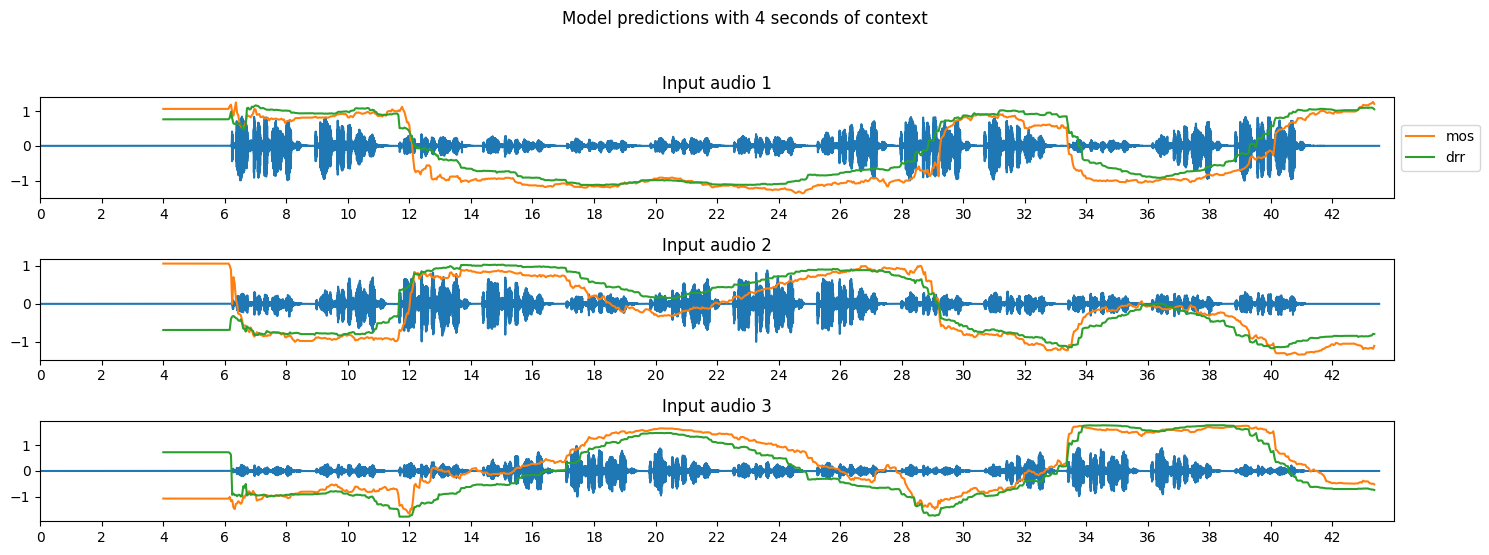}
    \caption{Multi-channel MOSRA predictions using a circular buffer of roughly 4 seconds of audio. The audio is recorded in a real room where the speaker is crossfaded between recording devices placed in three spatially disjoint locations. The model makes predictions on three channels, where the first two are repeated, e.g., the input to the model is channel [1,2,3,1,2]. Note that the scores are standardized across channels for each time step to aid interpretability.}
    \label{fig:realtime_preds}
\end{figure*}

\subsection{Speech quality}

Table \ref{table:MOS_RES} shows the performance of the baseline single-channel and multi-channel systems when trained on the simulated training data. The results show that in terms of MOS, the baseline single-channel system slightly but significantly outperforms the multi-channel system in Pearson correlation coefficient and RMSE metrics.

\subsection{Simulated dataset generalization}

To show the generalization capability of models trained on the simulated training data, we compare the original single-channel MOSRA model trained on human-labeled data \cite{hajal2022mosra} with the multi-channel model trained on our simulated data. To generate results with the multi-channel model on the single-channel data, we repeat the single-channel audio five times and take the average prediction at the output. We show results for in-distribution data, i.e., data generated in the same fashion where speech is convolved with an RIR and noise is added. We consider the NISQA TEST LIVETALK dataset as also in-distribution, as it contains only near and far field loud and quiet speech spoken in various noisy environments. The results in Table \ref{table:SIM_TO_REAL} show that for the in-distribution data, the model performs at least equally well as the original model. Out-of-distribution data is also shown in the form of the other NISQA TEST datasets, where the performance of the model trained on synthetic data is lower than the model trained on human-labeled data. These test data include distortions not seen during simulation (such as clipping, packet loss, and various codec distortions).

\begin{table}
\begin{tabular}{llll}
\toprule
\multicolumn{2}{l}{MOS on the test set}              &  & 95 \% CI \\
\midrule
\multirow{2}{*}{PCC}  & Single-channel (baseline)   & \textbf{0.97}  & [0.971, 0.973]         \\
                      & Multi-channel  &  0.96          & [0.957, 0.960]         \\

\midrule
\multirow{2}{*}{RMSE} & Single-channel (baseline)  & \textbf{0.21}  & [0.207, 0.213]         \\
                      & Multi-channel  & 0.28           & [0.277, 0.285]     \\
\bottomrule
\end{tabular}
\caption{PCC ($\uparrow$) and RMSE ($\downarrow$) of the single-channel baseline and multi-channel models on MOS labels using the simulated test set. The confidence intervals are obtained by bootstrapping with 1000 repetitions.}
\label{table:MOS_RES}
\end{table}

\begin{table}
\begin{tabular}{lllll}
\toprule

Pearson correlation       & Original  \cite{hajal2022mosra} & Simulated\\
after third-order mapping &  MOSRA &  data\\
\midrule

\multicolumn{3}{c}{In-distribution data} \\
\hdashline
    ReverbSpeechQuality\cite{reverbspeechquality} & 0.85 & \textbf{0.87}\\
    NISQA TEST LIVETALK\cite{Mittag_2021}     & 0.68 & \textbf{0.71} \\
\midrule

\multicolumn{3}{c}{Out-of-distribution data} \\
\hdashline
    NISQA TEST FOR\cite{Mittag_2021}   & \textbf{0.87} & 0.53 \\
    NISQA TEST P501\cite{Mittag_2021}   & \textbf{0.89} & 0.60\\
\bottomrule
\end{tabular}
\caption{PCC ($\uparrow$) after third order mapping on NISQA and internal test sets for the original MOSRA model trained on real human-labeled data vs. the multi-channel model trained on our simulated data.}
\label{table:SIM_TO_REAL}
\end{table}

\section{Discussion and Conclusion} 

Given that the architecture of our model allows for any length of audio input, the model could be used to give near real-time predictions on multiple channels using a circular buffer. Fig. \ref{fig:realtime_preds} shows the model predictions for MOS and DRR on three-channel audio where the speech is crossfaded between the different input channels. The figure shows that the model can detect the transitions of the speaker between channels and thus provides a proof of concept for the use of quality-based audio stream selection. Besides, Fig. \ref{fig:T60_DRR} underscores the limited correlation between DRR and the distance to speakers. This observation emphasizes the drawback of distance as a criterion for selecting the optimal device regarding audio quality, as it may overlook crucial factors such as DRR and T60. 

Our experimental findings confirm the advantages of a multi-channel model in predicting acoustic parameters, surpassing the single-channel counterpart. Notably, the multi-channel model improves prediction of STI, DRR, and C50, demonstrating a clear advantage over the single-channel model while utilizing 5$\times$ less computational resources per channel. Lower T60 prediction compared to the single-channel baseline might be caused by the model lacking capacity, as it is predicting five channels in parallel with roughly the same number of parameters and computational cost. However, we have not seen improvement in multi-channel quality (MOS) prediction, speculating the slightly worse performance against the single-channel version caused by missing multi-channel human speech quality labels.

On the other hand, the experiment on generalization shows that simulated data can allow for generalization to real data. Specifically, we observe that the model performs well on the NISQA TEST LIVETALK and ReverbSpeechQuality datasets but lacks performance on the two other NISQA datasets. Our conjecture is that the observed performance degradation is due to a mismatch in data distribution, as the high performance on the ReverbSpeechQuality could be attributed to the fact that data is generated in the same fashion. In line with previous findings on denoising~\cite{serrà2022universal}, we emphasize that future work should extend the types of degradations applied to speech to improve the generalization capability of MOS estimation models.

\vfill\pagebreak

\bibliographystyle{IEEEbib}
\bibliography{refs}

\end{document}